\documentclass[11pt]{elsart}
\usepackage{amsfonts}
\usepackage{slashbox}
\usepackage{amssymb,amsmath}
\usepackage{mathrsfs} 
\usepackage{graphicx}

\begin{document}

\begin{frontmatter}
\title{A note on revelation principle \\from an energy perspective}
\author{Haoyang Wu\corauthref{cor}}
\corauth[cor]{Corresponding author.} \ead{18621753457@163.com,
Tel: 86-18621753457}
\address{Wan-Dou-Miao Research Lab, Room 301, Building 3, 718 WuYi Road,\\
Shanghai, 200051, China.}

\begin{abstract}
The revelation principle has been known in the economics society for
decades. In this paper, I will investigate it from an energy
perspective, \emph{i.e.}, considering the energy consumed by agents and the
designer in participating a mechanism. The main result is that when the strategies of
agents are actions rather than messages, an additional energy condition should be
added to make the revelation principle hold in the real world.

JEL codes: D7
\end{abstract}


\end{frontmatter}

\section{Introduction}
The revelation principle is a fundamental theorem in economics
theory. According to the wide-spread textbook given by Mas-Colell,
Whinston and Green (Page 884, Line 24 \cite{MWG1995}): ``\emph{The
implication of the revelation principle is} ... \emph{to identify
the set of implementable social choice functions, we need only
identify those that are truthfully implementable}.''

So far, the revelation principle has been applied to many
disciplines such as auction, contract, the theory of incentives and
so on. If we move eyes from economics to physics, it is well-known
that the world is a physical world, doing any action requires
energy. In this paper, I will investigate the revelation principle
from an energy perspective, \emph{i.e.}, studying how much energy is
required for agents and the designer in participating a mechanism.
Section 2 and 3 are the main parts of this paper. Section 4 draws
conclusions. Related definitions and proofs are given in Appendix,
which are cited from Section 23.B and 23.D of MWG's textbook\cite{MWG1995}.

\section{Energy matrices}
Let us consider a setting with $I$ agents, indexed by $i=1,\cdots,I$
(page 858 \cite{MWG1995}). These agents make a collective choice
from some set $X$ of possible alternatives. Prior to the choice,
each agent $i$ privately observes his type $\theta_{i}$ that
determines his preferences. The set of possible types for agent $i$
is denoted as $\Theta_{i}$. The vector of agents' types
$\theta=(\theta_{1},\cdots,\theta_{I})$ is drawn from set
$\Theta=\Theta_{1}\times\cdots\times\Theta_{I}$ according to
probability density $\phi(\cdot)$. Each agent $i$'s Bernoulli
utility function when he is of type $\theta_{i}$ is
$u_{i}(x,\theta_{i})$. A mechanism
$\Gamma=(S_{1},\cdots,S_{I},g(\cdot))$ is a collection of $I$ sets
$S_{1}, \cdots, S_{I}$, each $S_{i}$ containing agent $i$'s possible
actions (or plans of action), and an outcome function
$g:S\rightarrow X$, where $S=S_{1}\times\cdots\times S_{I}$ (page
883, Line 7 \cite{MWG1995}).

At first sight, it is meaningless to discriminate whether the format of
agent $i$'s strategy is an action or a plan of action, since the two
formats of strategies correspond to the same results in the
traditional theory of mechanism design. However, as I will argue in the
following discussion, the two formats of strategies are different from an energy
perspective.

For any agent $i$, if his strategy $s_{i}(\cdot)$ is of an action
format, I denote by $E_{a}$ the energy required for agent $i$ to
to choose it (\emph{i.e.}, performing the action). Otherwise agent $i$'s
strategy $s_{i}(\cdot)$ is of a message format (\emph{i.e.}, a plan of
action), in this case I denote by $E_{m}$ the energy required for agent $i$
to choose it (\emph{i.e.}, selecting the message).

Generally speaking, an
action is laborious, to carry out it requires more energy; as a comparison, a
plan of action is a message, to select it requires less
energy. This is consistent to the common sense in the real world.
Therefore, it is natural to assume $E_{a}>E_{m}$. Note the private
type of agent $i$ can be reasonably represented as a message, because
agent $i$ can announce it to the designer. In addition, let
$E_{send}$ be the energy consumed in sending out a message
by agents, and $E_{g}$ be the energy consumed in performing the outcome
function $g(\cdot)$ by designer.

Now let us consider the revelation principle for Bayesian Nash
equilibrium given in Proposition 23.D.1 \cite{MWG1995}.
Suppose that there exists a mechanism
$\Gamma=(S_{1},\cdots,S_{I},g(\cdot))$ that implements the social
choice function $f(\cdot)$ in Bayesian Nash equilibrium, then
$f(\cdot)$ is truthfully implementable in Bayesian Nash equilibrium.
Let $\Gamma_{direct}=(\Theta_{1},\cdots,\Theta_{I},g(s^{*}(\cdot)))$
be the corresponding direct revelation mechanism.
According to the format of agents' strategies, an action or a plan of action,
there are two different cases:

\emph{Case 1}: The strategy is of a
message format (\emph{i.e.}, a plan of action).\\
1) Mechanism $\Gamma$: Given any $\theta\in\Theta$, each agent
$i$ selects the strategy $s^{*}_{i}(\theta_{i})$ and sends it to the
designer. Hence, the energy consumed by $I$ agents is
$I\cdot(E_{m}+E_{send})$. The designer receives $I$ messages and
performs the outcome function $g(\cdot)$. Hence, the energy consumed
by the designer is $E_{g}$.\\
2) Mechanism $\Gamma_{direct}$: Given any $\theta\in\Theta$,
each agent $i$ announces his/her type as a message to the designer. Hence,
the energy consumed by $I$ agents is $I\cdot E_{send}$. The designer
receives $I$ messages and performs the outcome function
$g(s^{*}(\cdot))$. Hence, the energy consumed by the designer is
$I\cdot E_{m}+E_{g}$.

\emph{Case 2}: The strategy is
of an action format. \\
1) Mechanism $\Gamma$: Given any $\theta\in\Theta$, each agent
$i$ performs his/her action $s^{*}_{i}(\theta_{i})$. Hence, the energy
consumed by $I$ agents is $I\cdot E_{a}$. The designer performs the
outcome function $g(\cdot)$. Hence, the energy consumed by the
designer is $E_{g}$.\\
2) Mechanism $\Gamma_{direct}$: Given any $\theta\in\Theta$,
each agent $i$ announces his/her type as a message to the designer. Hence,
the energy consumed by $I$ agents is $I\cdot E_{send}$. The designer
receives $I$ messages and perform the outcome function
$g(s^{*}(\cdot))$. Hence, the energy consumed by the designer is
$I\cdot E_{a}+E_{g}$.

The above-mentioned energy consumed in different cases can be represented
by an energy matrix in Table 1.

\emph{Table 1: An energy matrix of $I$ agents and the designer. The
first entry denotes the energy consumed by $I$ agents, and the
second stands for the energy consumed by the designer}.\\
\begin{tabular}{|c|c|c|}
\hline \backslashbox{Strategy format}{Mechanism} &
{$\Gamma$}&{$\Gamma_{direct}$}
 \\\hline A message & $[I\cdot(E_{m}+E_{send}), E_{g}]$ & $[I\cdot E_{send}, I\cdot E_{m}+E_{g}]$
\\ An action & $[I\cdot E_{a}, E_{g}]$ & $[I\cdot
E_{send}, I\cdot E_{a}+E_{g}]$
\\ \hline
\end{tabular}

Usually, $E_{a}$ is significant, as a comparison $E_{m}$, $E_{g}$ and $E_{send}$ are small. Suppose
$E_{m}$, $E_{g}$ and $E_{send}$ can
be neglected, then Table 1 is reduced to Table 2:

\emph{Table 2: A simplified energy matrix of $I$ agents and the designer}.\\
\begin{tabular}{|c|c|c|}
\hline \backslashbox{Strategy format}{Mechanism} & {
$\Gamma$}&{$\Gamma_{direct}$}
 \\\hline A message & $[0, 0]$ & $[0, 0]$
\\ An action & $[I\cdot E_{a}, 0]$ & $[0, I\cdot E_{a}]$
\\ \hline
\end{tabular}

In terms of computer science, when agents' strategies are actions
instead of messages (\emph{i.e.}, plans of action), the complexity of the energy consumed by
the designer in $\Gamma_{direct}$ is $\mathcal{O}(I)$, which cannot
be neglected. Therefore, in order to make the direct revelation
mechanism $\Gamma_{direct}$ work, an additional energy condition should be
added as follows:\\
\emph{Energy condition}: The designer possesses enough energy, at least the sum
of energy that all agents would consume when they participate the
original indirect mechanism $\Gamma$.

\section{Discussions}
In this section, I will analyse two problems facing the
designer when agents' strategies are of an action format:\\
1) In the direct mechanism $\Gamma_{direct}$, \emph{does the
designer possess enough energy to carry out all actions that would
be consumed by agents in the original indirect mechanism $\Gamma$?}
(Generally speaking, there are many factors that may be relevant to
agents' actions, \emph{e.g.}, energy, skill, quality, etc. For simplicity,
here I only consider one indispensable factor, \emph{i.e.}, the energy
required to carry out an action.)

According to Page 378, the 9th line to the last \cite{Serrano2004},
``... \emph{the mechanism designer is always at an informational
disadvantage with respect to the agents, who, as a collective
entity, know more about the true environment that does the
designer}''. Similar to this idea, it looks somewhat ``unreasonable''
to assume that the designer is at an energy advantage with respect
to the agents, \emph{i.e.}, the designer possesses enough energy that is
not less than the sum of all agents' energy.

As shown in Table 2, the energy condition is very weak when the
strategies of agents are of a message format. However, when the
strategies of agents are of an action format, the energy condition
may be restrictive. The designer cannot take it for granted that he
always has enough energy.
When the power of the designer is restricted such that the energy
condition does not hold, the revelation principle will not hold.

2) Furthermore, even if the energy condition is satisfied, there
still exists another problem facing the designer. As shown in Table
2, when the designer chooses the indirect mechanism $\Gamma$, he
almost spends zero energy; but if the designer chooses the direct
mechanism $\Gamma_{direct}$, he has to spend $I\cdot E_{a}$ energy
to make $\Gamma_{direct}$ work. Note that in the theory of mechanism
design, the designer only care whether and how the social choice
function $f(\cdot)$ can be implemented. Since $\Gamma$ and
$\Gamma_{direct}$ implement the same $f(\cdot)$ in Bayesian Nash
equilibrium, then \emph{why does the designer have incentives to
work harder, i.e., to be willing to choose $\Gamma_{direct}$ instead
of $\Gamma$?}

\section{Conclusion}
In this paper, two main results are yielded: \\
1) If the strategies of
agents are of an action format, then an energy condition should be
added to make the revelation principle hold in the real world.
Furthermore, it is questionable to say that the designer has
incentives to work harder by choosing a direct mechanism, but
finally implements the same social choice function as he would
implemented easier by choosing an indirect mechanism. Hence, the revelation principle may
be not proper when agent's strategies are of an action format.

2) If the strategies of agents
are of a message format (\emph{i.e.}, plans of actions), then there is no problem in the
traditional framework of revelation principle. Note: this result
holds under the assumption that $E_{m}$, $E_{g}$ and $E_{send}$ can
be neglected.\\

\section*{Acknowledgments}

The author gratefully acknowledges helpful conversations with Dr.
Hongtao Zhang. The author is also very grateful to Ms. Fang Chen,
Hanyue Wu (\emph{Apple}), Hanxing Wu (\emph{Lily}) and Hanchen Wu
(\emph{Cindy}) for their great support.

\section*{Appendix: Definitions in Section 23.B and 23.D
[1]} \textbf{Definition 23.B.1}: A social choice function is a
function $f:\Theta_{1}\times\cdots\times\Theta_{I}\rightarrow X$
that, for each possible profile of the agents' types
$(\theta_{1},\cdots,\theta_{I})$, assigns a collective choice
$f(\theta_{1},\cdots,\theta_{I})\in X$.

\textbf{Definition 23.B.3}: A mechanism
$\Gamma=(S_{1},\cdots,S_{I},g(\cdot))$ is a collection of $I$
strategy sets $S_{1},\cdots,S_{I}$ and an outcome function
$g:S_{1}\times\cdots\times S_{I}\rightarrow X$.

\textbf{Definition 23.B.5}: A direct revelation mechanism is a
mechanism in which $S_{i}=\Theta_{i}$ for all $i$ and
$g(\theta)=f(\theta)$ for all
$\theta\in\Theta_{1}\times\cdots\times\Theta_{I}$.

\textbf{Definition 23.D.1}: The strategy profile
$s^{*}(\cdot)=(s^{*}_{1}(\cdot),\cdots,s^{*}_{I}(\cdot))$ is a
\emph{Bayesian Nash equilibrium} of mechanism
$\Gamma=(S_{1},\cdots,S_{I},g(\cdot))$ if, for all $i$ and all
$\theta_{i}\in\Theta_{i}$,
\begin{equation*}
  E_{\theta_{-i}}[u_{i}(g(s^{*}_{i}(\theta_{i}),s^{*}_{-i}(\theta_{-i})),\theta_{i})|\theta_{i}]
  \geq
  E_{\theta_{-i}}[u_{i}(g(\hat{s}_{i},s^{*}_{-i}(\theta_{-i})),\theta_{i})|\theta_{i}]
\end{equation*}
for all $\hat{s}_{i}\in S_{i}$.

\textbf{Definition 23.D.2}: The mechanism
$\Gamma=(S_{1},\cdots,S_{I},g(\cdot))$ implements the social choice
function $f(\cdot)$ in Bayesian Nash equilibrium if there is a
Bayesian Nash equilibrium of $\Gamma$,
$s^{*}(\cdot)=(s^{*}_{1}(\cdot),\cdots,s^{*}_{I}(\cdot))$, such that
$g(s^{*}(\theta))=f(\theta)$ for all $\theta\in\Theta$.

\textbf{Definition 23.D.3}: The social choice function $f(\cdot)$ is
truthfully implementable in Bayesian Nash equilibrium if
$s^{*}_{i}(\theta_{i})=\theta_{i}$ (for all
$\theta_{i}\in\Theta_{i}$ and $i=1,\cdots,I$) is a Bayesian Nash
equilibrium of the direct revelation mechanism
$\Gamma=(\Theta_{1},\cdots,\Theta_{I},f(\cdot))$. That is, if for
all $i=1,\cdots,I$ and all $\theta_{i}\in\Theta_{i}$,
\begin{equation*}
  E_{\theta_{-i}}[u_{i}(f(\theta_{i},\theta_{-i})),\theta_{i})|\theta_{i}]
  \geq
  E_{\theta_{-i}}[u_{i}(f(\hat{\theta}_{i},\theta_{-i}),\theta_{i})|\theta_{i}],\quad\quad
  (23.D.1)
\end{equation*}
for all $\hat{\theta}_{i}\in \Theta_{i}$.

\textbf{Proposition 23.D.1} (\emph{The Revelation Principle for
Bayesian Nash Equilibrium}) Suppose that there exists a mechanism
$\Gamma=(S_{1},\cdots,S_{I},g(\cdot))$ that implements the social
choice function $f(\cdot)$ in Bayesian Nash equilibrium. Then
$f(\cdot)$ is truthfully implementable in Bayesian Nash equilibrium.

\end{document}